# Formation of structure for gas-liquid, non-equilibrium polymer media

T.V. Savenkova[1], M.A. Taleisnik[1], A.R. Karimov[2,3], T.V. Gerasimov[1], A.M. Bulygin[3], S. A. Terekhov[3]

[1] All-Russian Scientific Research Institute of Confectionery Industry - Branch of V.M. Gorbatov Federal Research Center for Food Systems RAS, Moscow, Russia

[2] Joint Institute for High Temperatures RAS, Moscow, Russia

[3] National Research Nuclear University MEPhI, Moscow, Russia

This paper examines the mechanisms of destruction and synthesis for macromolecules, which may be propelled by external acoustic fields in the flows of polymeric liquids containing a large number of gas bubbles. The dynamics of these bubbles is assumed to govern by changing the flow geometry and exciting sound oscillations in the flow. Mechanically-induced kinetic changes in macromolecules (destruction and synthesis of polymer chains) will occur when the bubbles collapse.

## 1. Introduction

From the formal point of view, the existing methods for food recycling may be classified by mechanical, thermal, biotechnological, biochemical and chemical properties [1]. The type of the energy supply, energy redistribution and, also, energy dissipation in the technological system determines such classification of technological processes. This means that the incoming energy fluxes are applied for creation a new formation in food (for example, structural transformation in a medium occurs in appropriate kinetic reactions) and for useless losses. These two factors determine the final product quality despite of the previous formal separation.

These features are particularly manifested in the traditional technologies of production of confectionery masses (see, for example, [1-4]) which are produced in the local equilibrium condition. It implies that such technological processes are so slow that at each moment of time some equilibrium macroscopic state is established in the system, and we can describe this state with such macroscopic parameters as a density or molar concentrations for multiphase flow, a temperature. In this case, the impact on the medium begins on molecular and finishes on macroscopic spatial scale in all technological volume. In addition, the kinetic reactions last until they reach the appropriate local equilibrium state values.

In an ideal case, we would like to escape the useless energy losses. Moreover, the biological properties of the processed medium may get worse because of an excessive heat treatment. In contrast to the routine local equilibrium approach, we would like to make locally changes in structure of the treated medium, which, however, has drastically to change its physical and chemical properties in whole.

To our opinion, such issue can be solved only in the framework of nonequilibrium approach (for example, see [5,6]). In the present paper we are going to discuss the possible way to this issue taking into account the features of structure formation of the used polymer media.

**2. Special structure properties of polymer media**

As a rule, the confectionery media have used in the form of solutions and melts, emulsions, invert and concentrated sugar syrups. They contain a significant amount of large macromolecules (polymers). For example, invert syrup contains not less than 65% of dry substances, including sucrose, fructose and glucose, and the emulsion also contains dairy and egg products, fats [7]. As a typical example of structure for such macromolecules, the schematic of sucrose macromolecules is presented in Fig. 1; they are built from repetitive building blocks — monomers, connected by chemical bonds in a single chain.

As is shown from this figure, all macromolecules are characterized by a developed internal structure and the presence of various degrees of freedom, which determine the specific macroscopic properties of polymer liquids. This point allows us to classify these media as a class of the so-called non-Newtonian fluids whose rheological properties are fundamentally different from those of simple liquids [6,9]. For example, such media are characterized by viscoelastic properties, namely, if we stretch the polymer melt and cut it in the middle, the halves will be reduced to almost the original state. The manifestation of such a high degree of elasticity means that during the flow of polymers there occurs a forced change in the arrangement of atoms (such a process is called conformation) and a change in the number of cohesions between them. In this case, the monomeric units are bound in long chains and all atoms of macromolecules are displaced in one direction, i.e. they lack an independent translational movement. These properties can cause the accumulation of energy in individual oscillation modes, which are essentially individual monomers (see Fig. 1).

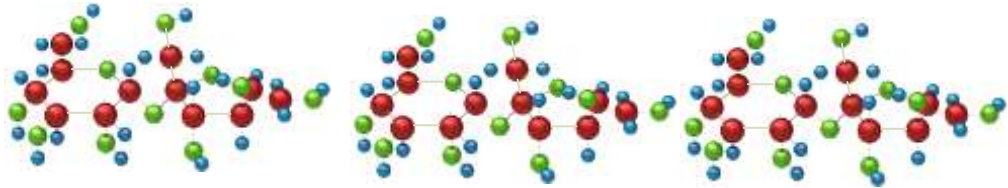

Fig. 1. The peculiarities of polymer's structures: sucrose molecule (glucose and fructose) [7]

One should be borne in mind that the number of monomers determines the permissible number of freedom degrees. At the same time, each degree of freedom carries a certain energy and ensures the spatial arrangement of monomers, i.e. the stored energy may be spent on the formation of polymers with various configurations [6]: chains can roll into tangles and vice versa, and under certain conditions such supramolecular formations can stretch into chains (see Fig. 2).

At the same time, these processes bring about a change in the medium density with the formation of voids which under certain conditions can become nuclei containing the gas phase and change over time (see Fig. 3) [6]. It means that the medium inevitably becomes compressible and, accordingly, compressibility is a necessary condition for further the growth of voids (see Fig. 2) [7,11]. A typical example of such behavior is the dynamics of emulsions in which the density in the process of mechanical action decreases with respect to time [7]. Thus, one can say that a fundamental feature of polymeric liquids is the presence of a huge number of cavitation nuclei that can under certain conditions become the macroscopic gas bubbles.

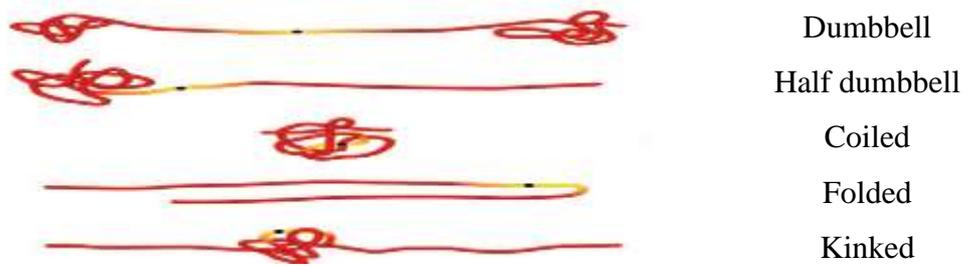

| | |
|---|---|
| | Dumbbell |
| | Half dumbbell |
| | Coiled |
| | Folded |
| | Kinked |

Рис. 2. Schematics of different possible polymer conformations [6]

## 3. The cavitation bubble dynamics in non-equilibrium environment

It is possible to influence the dynamics of these nuclei from a macroscopic level to a microscopic level on account of the excitation of acoustic waves in a flow of polymer liquid with a wavelength commensurable with the characteristic size of the polymer macromolecule [5,6]. In this case, a local supply of energy to individual sections of the medium will occur.

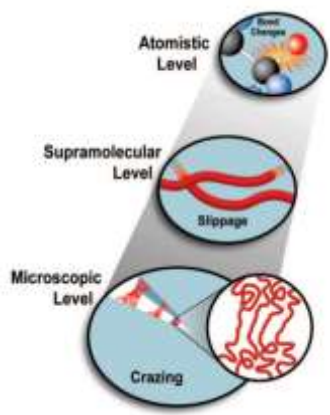

1. Atomic level – changes in chemical bonding and conformational changes.
2. The supermolecular level – the slippage of polymer chains, as a reaction to external influence and deformation.
3. Microscopic level – the formation of voids, cavitation nuclei, cracks and large-scale viscoelastic deformations

Fig. 3. Hierarchical levels of mechanical and chemical changes [6]

Acoustic wave in the air:
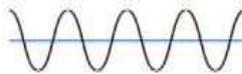

Acoustic wave in the liquid:
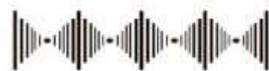

Development of cavitation nuclei:
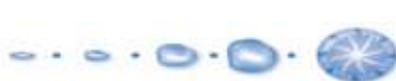

Cavitation dynamics:
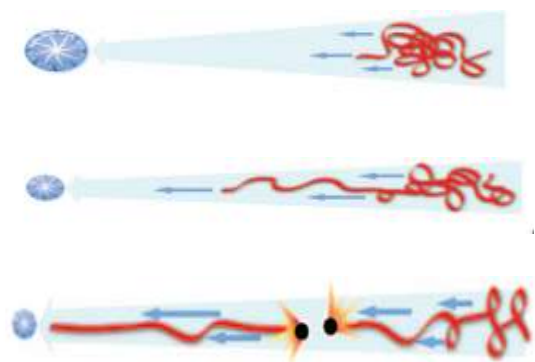

Fig. 4. Acoustic method to control the dynamics of bubbles [6,1]

This selective effect may initiate various kinetic reactions in these regions, changing the structure of the medium. In particular, it will lead to the development of individual microscopic voids, and filling them with gas contained in the medium (Fig. 4) [6,11]. In simple liquids, for example, in water, the development of gas bubbles can be accompanied by coagulation of the bubbles and their emergence from the medium under the action of Archimedean force [12]. In this case, one should take into account the fundamental difference from the dynamics of bubbles in simple liquids, caused to the structural features of polymer macromolecules. The molecules located at the interface with a gas can be considered at the persistent length as dipoles, at the ends of which the charges of different signs are concentrated [12]. It can be assumed that the interacting molecules are stretched in the same direction and the signs of their charges on each side are also the same. As a result, a membrane effect occurs at the interface, which prevents the transit of certain molecules and ions. In this case, the border has a so-called amphiphilic structure that determines the basic properties of cell membranes. Moreover, when two bubbles contact with such boundary, the amphiphilic structure of the skin at the boundary of each bubble prevents them from merging. The interruption of a large number of bubbles leads to a formation of the foam, where neighboring bubbles are separated by a skin. This mechanism determines the structure of many food polymeric masses.

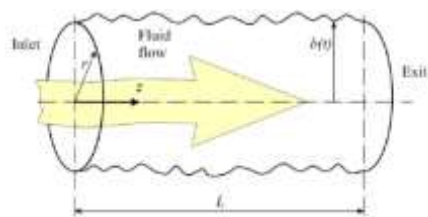

Fig. 5. The flow scheme [11]

In contrast to simple liquids in non-equilibrium polymer liquids coagulation of bubbles is impossible on account of the amphiphilic effect. In this case, the change in dynamic pressure on the surface of the bubbles, on account of the acoustic effect, with a decrease in the total hydrodynamic pressure in the flow region containing the growing bubbles, leads to an increase in the size of the bubble and surface energy. This process arises due to, for example, a change in the configuration limiting the flow of the surface (see Fig. 5), when, with increasing velocity, a drop in hydrodynamic pressure occurs. However, the acceleration of the flow can be replaced by its inhibition, when the kinetic energy of the liquid turns into a thermal form. As a result, the pressure increase will occur, leading to a collapse of the bubble with the release of stored energy in the so-called singular point, which can lead to the rupture of polymer chain [5,11].

## 4. Kinetics and hydrodynamic singularities

At this singular point, there will be an instantaneous increase in the concentration of matter and energy that will create favorable conditions for accelerating kinetic reactions which in traditional food technologies take place at the elevated temperatures throughout the technological volume or cannot be initialized at all. For example, the energy released during the collapse of a bubble may be sufficient for the excitation, ionization and dissociation of water molecules and gases inside the cavitation cavity [13]. In this case, we may venture a guess that these particles colliding with fragments of broken polymer macromolecules will become part of these fragments. Thus, fundamentally new macromolecular structures can arise.

In order to show the influence of singular dynamics on the chemical kinetics we consider the one-dimensional flow [14-16]. For a one-dimensional flow with first-order reactions (noting that such reactions can simulate the splitting of the polymer chain), the density distribution in the Lagrangian coordinates (x, t) has the form [16]:

$$n(x,t) = \frac{\gamma n_0 e^{-kt}}{\gamma + (1-e^{-\gamma t})\frac{dV_0}{dx}}, \qquad (1)$$

where $\gamma$ is the collision frequency, $k$ is the chemical reaction constant, $n_0$ is the initial density of the flow, $V_0$ is the initial velocity of the flow. As is seen from the present relation, for $V_0' < 0$ always there is a moment of time $t = t_*$, when the denominator vanishes, i.e. density increases indefinitely at this moment for some coordinate x= $x_*$. Since the rate of the corresponding kinetic reaction is proportional to the density and the constant $k$, depending on temperature, the reaction rate will also increase indefinitely in the vicinity of x= $x_*$ at a given point in time.

Moreover, there are initial conditions for which one can come about the multiple collapses formation. We display such behavior in the limit $\gamma \to 0$, $k \to 0$ for the velocity initial distribution [14]:

$$V_0(x) = x + B sin\left(\frac{x}{L}\right), \qquad (2)$$

and the initial Gaussian profile of density

$$n_0(x) = \exp\left(-\frac{x^2}{2}\right), \qquad (3)$$

where $B$ and L are some positive constants. In the case B/L>>1 there appears a distribution of density depicted in Fig. 6.

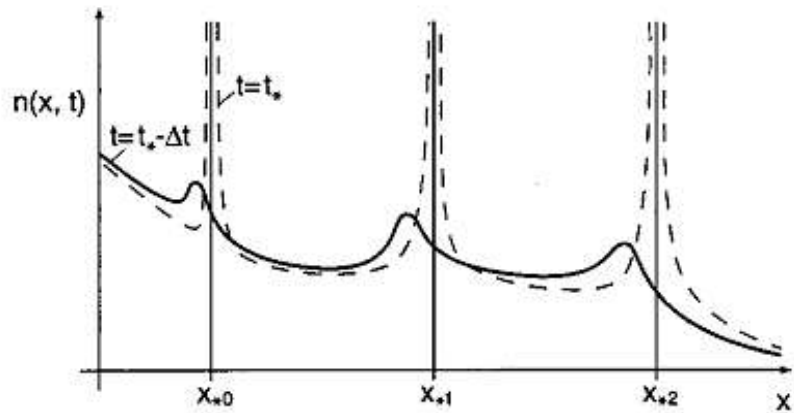

Рис. 6. The formation of density profile with multiple collapses [14]

Proceeding from this simple model one can assume that for an axisymmetric flow with appropriate velocity field, in which acoustic oscillations are excited one can expect the formation of multiple density collapses in collapsing bubbles [6, 11, 17]. Thus, the excitation of acoustic oscillations in the flow of polymer liquid, which in a given way is either accelerated or slowed down, can provide a local effect on the structure of the polymer medium.

## 5. Discussion

At present, technically, the combination of hydrodynamic and acoustic effects, under which conditions the passage of physicochemical processes is initiated, is implemented on a laboratory cavitation facility, ensuring maximum energy storage and its release with an intense technological effect (see Fig. 7) [1,17].

As an object of research, one of the simplest polymers was chosen - invert syrup, since it will most clearly appear the basic properties of polymers. On the other hand, it is one of the main semi-finished products of confectionery production consisting of glucose and fructose macromolecules.

As a result of the studies carried out on the invert syrup at the facility under conditions of cavitation impact, a well-observed change in the structure of food materials of the nanometer range was obtained for the first time (Fig. 8).

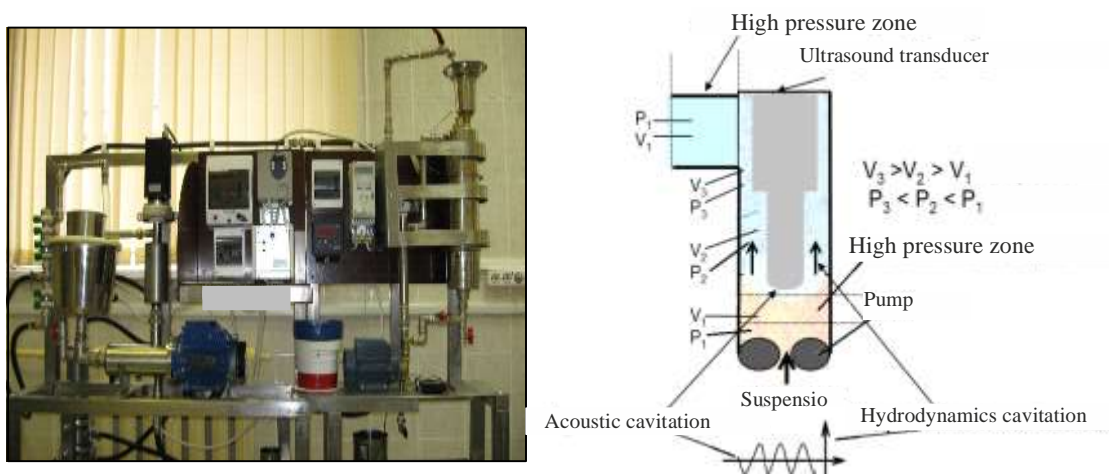

Fig. 7. Laboratory cavitation facility [1]; here, the flow is realized the velocity distribution $V_3 > V_2 > V_1$ and the pressure distribution $P_3 < P_2 < P_1$

| without cavitation | with cavitation |
|---|---|
| 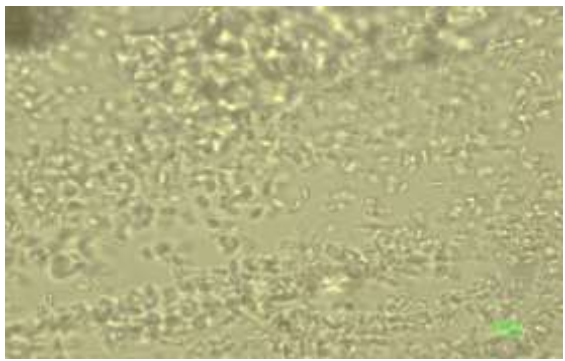 | 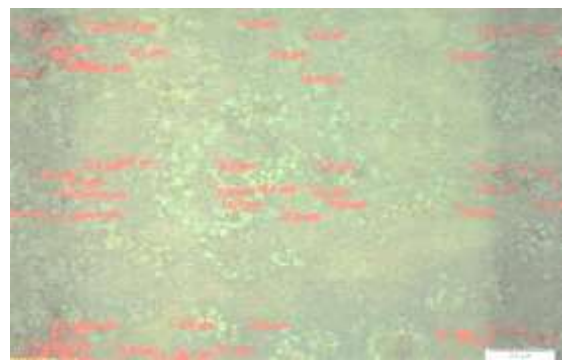 |
| Particle size d = 2000 - 3000 nm multiplication 500 | Particle size d = 72 - 230 nm multiplication 1000 |

Fig. 8. Micro photo of the invert syrup

Thus, the concept of controlling kinetic transformations in polymeric media has been proposed and justified. This creates the basis for developing a tool for changing the structure and properties of individual macromolecules in polymeric media, including food systems [1,17,18].